\documentclass[aps,twocolumn,preprintnumbers,showpacs,showkeys,nofootinbib%
]{revtex4}
\usepackage{epsfig}
\usepackage{amssymb,amsmath,amsfonts,amsthm,graphicx,psfrag}
\graphicspath{{./Figures/}}                 
\newcommand{\noi}{\noindent}
\newcommand{\beq}{\begin{equation}}
\newcommand{\eeq}{\end{equation}}
\newcommand{\bea}{\begin{eqnarray}}
\newcommand{\eea}{\end{eqnarray}}
\newcommand{\Fig}[1]{Fig.~\ref{#1}}
\newcommand{\Tab}[1]{Table~\ref{#1}}
\newcommand{\Sec}[1]{Section~\ref{#1}}
\newcommand{\Eq}[1]{Eq.~(\ref{#1})}

\def\ds{\displaystyle}
\def\U{{U_{x,\mu}}}
\def\Ud{{U^\dagger_{x,\mu}}}

\newcommand{\tr}{\operatorname{Tr}}

\newcommand{\bc}{{\it bc~}}
\newcommand{\fc}{{\it fc~}}
\begin{document}

\title{Lattice Gluon Propagators in $3d$ \, $SU(2)$ Theory and Effects of Gribov
Copies }

\author{V.~G.~Bornyakov}
\affiliation{Institute for High Energy Physics, 142281, Protvino, Russia \\
and Institute of Theoretical and Experimental Physics, 117259 Moscow, Russia}

\author{V.~K.~Mitrjushkin}
\affiliation{Joint Institute for Nuclear Research, 141980 Dubna, Russia \\
and Institute of Theoretical and Experimental Physics, 117259 Moscow, Russia}

\author{R.~N.~Rogalyov}
\affiliation{Institute for High Energy Physics, 142281, Protvino, Russia }

\date{06.11.2012}

\begin{abstract}
Infrared behavior of the Landau gauge gluon propagators is studied
numerically in the 3d $SU(2)$ gauge theory on the lattice.  A special
accent is made on the study of Gribov copy effect.  For this study
we employ an efficient gauge-fixing algorithm and a large number of
gauge copies (up to 280 copies per configuration).  It is shown that,
in the deep infrared region, the Gribov copy effects  are very significant.
Also we show that, in the infinite-volume limit, the zero-momentum value
of the propagator does not vanish.
\end{abstract}



\maketitle

\section{Introduction}
\label{sec:introduction}

The nonperturbative (first principle) numerical
computation of the field propagators is important for various reasons.
There are scenarios of confinement based on infrared behavior of the
gauge dependent propagators.
In particular, in the Gribov-Zwanziger
(GZ) confinement scenario \cite{Gribov:1977wm, Zwanziger:1991gz} the
Landau-gauge gluon propagator $D(p)$ at infinite volume is expected to
vanish in the infrared (IR) limit $p\to 0$.  On the other hand, a refined
Gribov-Zwanziger (RGZ) scenario proposed recently \cite{Dudal:2007cw,
Dudal:2008sp,Dudal:2008rm} allows a finite nonzero value of $D(0)$ .
Another reason is that the nonpertubative lattice calculations are
necessary to check the results obtained by analytical methods, e.g.,
the Dyson-Schwinger equations (DSE) approach which uses truncations
of the infinite set of equations.  The DSE scaling solution
predicts that the propagator tends to zero in the zero-momentum
limit \cite{vonSmekal:1997is,Alkofer:2000wg} in accordance
with the RGZ-scenario.  At the same time, the DSE decoupling solution
\cite{Cornwall:1981zr,Fischer:2008uz,Aguilar:2008xm,Boucaud:2008ji} allows a finite
nonzero value of $D(0)$ in conformity with RGZ-scenario.

The $3d$  $SU(2)$ theory can serve as a useful testground to verify
these predictions in a simplified, in comparison with the $4d$ case,
setting.  Furthermore, $3d$ theory is of interest for the studies of
the high-temperature limit of the $4d$ theory.

The most theoretically attractive definition of the Landau gauge is to
choose for every gauge orbit a representative from the fundamental modular
region \cite{SemenovTyanShanskii}, i.e. the {\it absolute} maximum of the gauge fixing functional
$F(U)$ (see the definition in Section \ref{sec:definitions}.).

The arguments in favor of this choice are the following :
\noi \hspace{1mm} {\bf a)} a consistent non-perturbative gauge
fixing procedure proposed by Parrinello--Jona-Lasinio and Zwanziger
(PJLZ-approach) \cite{Parrinello:1990pm, Zwanziger:1990tn} presumes
that the choice of a unique representative of the gauge orbit should be
through the {\it global} extremum of the chosen gauge fixing functional;
\noi \hspace{1mm} {\bf b)} in the case of pure gauge $U(1)$ theory in the
weak coupling (Coulomb) phase some of the gauge copies produce a photon
propagator with a decay behavior inconsistent with the expected zero mass
behavior \cite{Nakamura:1991ww,Bornyakov:1993yy,Mitrjushkin:1996fw}.
The choice of the global extremum permits to obtain the physical -
massless - photon propagator.
It should be noted that, for practical purposes, it is sufficient to
approach the global maximum close enough so that the systematic errors
due to nonideal gauge fixing (because of, e.g., Gribov copy effects)
are  of the same  magnitude as statistical errors.  We follow here this
strategy which has been checked already in many papers on $4d$ theory
studies for both $SU(2)$ \cite{Bogolubsky:2005wf, Bogolubsky:2007bw,
Bornyakov:2008yx, Bornyakov:2009ug, Bornyakov:2010nc} and $SU(3)$
\cite{Bogolubsky:2007bw,Bogolubsky:2009dc,Aouane:2011fv, Bornyakov:2011jm}
gauge groups.

The three-dimensional $SU(2)$ theory has been recently studied in
\cite{Cucchieri:2003di, Cucchieri:2004mf,
Cucchieri:2007md,Cucchieri:2007ta, Maas:2008ri,Cucchieri:2011ig}.
The evidence has been presented that the propagator has a maximum
at momenta about 350 MeV and that $D(0)$ does not converge to zero
in the infinite-volume limit.  The problem of Gribov-copy
effects was addressed in \cite{Maas:2008ri}. The Gribov-copy
effects for the gluon propagator were found for small
momenta. We will show in this paper that these effects were
underestimated.

As compared to the previous version of this article,
we take into account the lattices $80^3$ and $96^3$
and obtain better statistics  on lattices of smaller size.
For this reason, we make the respective improvements in our plots, and
draw more definite conclusions.

In \Sec{sec:definitions} we introduce the quantities to be computed.
In \Sec{sec:details} some details of our simulations are given.
In \Sec{sec:gribov} we discuss the effect of improved gauge
fixing and present our numerical results.
Conclusions are drawn in \Sec{sec:conclusions}.

\section{The gluon propagator: definitions}
\label{sec:definitions}

We consider cubic $L\times L \times L$ lattice $\Lambda$ with spacing $a$.
To generate Monte Carlo ensembles of thermalized configurations we use
the standard Wilson action

\beq S  = \beta \sum_{x,\mu >\nu} \left[ 1 -\frac{1}{2}~\tr
\Bigl(U_{x\mu}U_{x+\hat \mu a;\nu} U_{x+\hat \nu
a;\mu}^{\dagger}U_{x\nu}^{\dagger} \Bigr)\right]\,,
\label{eq:action} \eeq

\noindent where $\beta = 4/g^2a$,  $\hat \mu$ is a  vector of unit
length along the $\mu$th coordinate axis and $g$ denotes
dimensionful bare coupling.  $U_{x\mu} \in SU(2)$ are the link
variables which transform under local gauge transformations $g_x$
as follows:

\beq U_{x\mu} \stackrel{g}{\mapsto} U_{x\mu}^{g} = g_x^{\dagger}
U_{x\mu} g_{x+\hat \mu a} \,, \qquad g_x \in SU(2) \,.
\label{eq:gaugetrafo} \eeq

\noi We study  the gluon propagator

\bea
D^{bc}_{\mu\nu}(q) &=& {a^3 \over L^3} \sum_{x, y \in \Lambda} \exp \left(iqx+{ia\over 2}\, q(\hat \mu - \hat\nu)\right)\\ \nonumber
&& \langle A^b_\mu(x+y+{\hat \mu a\over 2}) A^c_\nu(y+{\hat \nu a\over 2}) \rangle, \nonumber
\eea

\noindent where the vector potentials are defined as follows
\cite{Mandula:1987rh}:

\beq
 A_\mu\left(x+{\hat \mu a\over 2}\right) \equiv \sum_{b=1}^3 A^b_\mu {\sigma^b\over 2} =
{i\over ga}\big( \U - \Ud \big),
\eeq

\noindent and the momenta $q_\mu$ take the values $q_\mu =
2\pi n_\mu/aL$, where $n_\mu$ runs over integers in the range
$-L/2 \leq n_\mu < L/2$.
The gluon propagator can be represented in the form

\begin{displaymath}
D^{bc}_{\mu\nu}(q) = \left\{ \begin{array}{l}
\delta^{bc} \delta_{\mu\nu} \bar D(0), \qquad\qquad\qquad p=0;  \\
\delta^{bc} \left(\ds \delta_{\mu\nu} - {p_\mu p_\nu \over p^2 }\right)\; \bar D(p) , \qquad p\neq 0,
\end{array} \right.
\end{displaymath}

\noindent where $\ds ~p_\mu = {2\over a} \sin {q_\mu a\over 2}~$ and
$ ~p^2 = \sum_{\mu=1}^3 p_\mu^2$.
When $p\neq 0$, we arrive at
\beq\label{eq:QuantityUnderStudyDef}
\bar D(p)= {1\over 6}\, {1 \over ( La)^3}\  \sum_{\mu=1}^{3} \sum_{b=1}^3
\langle \tilde A_\mu^b(q)\, \tilde A_\mu^b(-q) \rangle,
\eeq
where
\beq
\tilde A_\mu^b(q) = a^3 \sum_{x\in \Lambda} A_\mu^b\left(x+{\hat \mu a\over 2}\right)
\exp\left(\ iq(x+{\hat \mu a\over 2}) \right),
\eeq
and the zero-momentum propagator has the form
\beq
\bar D(0)= {1\over 9}\, {1 \over ( La)^3}\  \sum_{\mu=1}^{3} \sum_{b=1}^3
\langle \tilde A_\mu^b(0)\, \tilde A_\mu^b(0)   \rangle.
\eeq





\noi In the weak-coupling infinite-volume limit, $\ds \bar D(p) =
{1\over p^2}$.  In this work, we  deal with the quantity (referred to
as the gluon propagator) \footnote{Note that the function $\bar D(p)$
(not $D(p)$!) is normalized similar to the analogous function used in
\cite{Cucchieri:2007md}  and \cite{Maas:2008ri}.}

\beq \label{eq:OurPropDef}
D(p)= \bar D(p)/\beta;
\eeq

In lattice gauge theory the usual choice of the Landau gauge
condition is~\cite{Mandula:1987rh} \beq (\partial A)(x) = {1\over
a}\ \sum_{\mu=1}^3 \left( A_\mu(x+{\hat\mu a\over 2})
  - A_\mu(x-{\hat\mu a\over 2}) \right)  = 0 \,
\label{eq:diff_gaugecondition}
\eeq
which is equivalent to finding a local extremum of the gauge functional
\beq
F_U(g) = ~\frac{1}{3L^3}\sum_{x\mu}~\frac{1}{2}~\tr~U^{g}_{x\mu}
\label{eq:gaugefunctional}
\eeq
with respect to gauge transformations $~g_x$.
The manifold consisting of Gribov copies providing local
maxima of the functional (\ref{eq:gaugefunctional}) and a semi-positive
Faddeev-Popov operator is termed the {\it Gribov region} $~\Omega$, while
that of the global maxima is termed the {\it fundamental modular region} (FMR)
$~\Gamma \subset \Omega$.  Our gauge-fixing procedure is aimed to approach
$~\Gamma$.

\section{Details of the simulation}
\label{sec:details}

In this work, we are going to demonstrate that the gluon propagator
in the deep infrared region can be reliably evaluated only when
the effects of Gribov copies are properly taken into account.
We make simulations at a fixed gauge coupling on
various lattices using gauge-fixing algorithm which was already
succesfully employed in the $4d$ theory at both zero \cite{Bornyakov:2008yx,Bornyakov:2009ug} and
nonzero \cite{Bornyakov:2010nc,Bornyakov:2011jm} temperature. There are three main ingredients of
this algorithm: powerful simulated annealing algorithm, which proved to be efficient in solving various
optimization problems; the flip transformation of gauge fields, which was
used to decrease both the Gribov-copy and finite-volume effects
\cite{Bornyakov:2008yx,Bornyakov:2009ug,Bornyakov:2010nc};
simulation of a large number of gauge copies for each flip sector in order
to further decrease the effects of Gribov copies.

Most of our Monte Carlo simulations has been performed at $\beta = 4.24$
for various lattice sizes $L$.  To study lattice spacing dependence of
the Gribov copies effects we have made also simulations on lattices
with  smaller lattice spacing ($a=0.94$ fm,~~$\beta = 7.09$) and
physical lattice size 6.03 fm ($L=64$) matching that of $L=36$ lattice
at $\beta=4.24$.

Consecutive configurations (considered to be statistically independent)
were separated by 200 sweeps, each sweep consisting of one local heatbath
update followed by $15$ microcanonical updates. In \Tab{tab:data_sets}
we provide the full information about the field ensembles used in this
investigation.

\begin{table}[h]
\begin{center}
\vspace*{0.2cm}
\begin{tabular}{|c|c|c|c|c|} \hline
 $~L~$ & $~n_{meas}~$ & $n_{copy}$ & $aL$[fm] &  $p_{min}$[GeV]
\\ \hline\hline
  32   & 800  &  96 &   5.38  &  0.230   \\
  36   & 1037 & 160 &   6.05  &  0.204   \\
  40   & 1032 & 160 &   6.73  &  0.184   \\
  44   & 717  & 160 &   7.39  &  0.167   \\
  48   & 1425 & 160 &   8.08  &  0.153   \\
  52   & 1085 & 160 &   8.74  &  0.141   \\
  56   & 796  & 160 &   9.43  &  0.131   \\
  64   & 709  & 160 &   10.8  &  0.115   \\
  72   & 910  & 280 &   12.1  &  0.102   \\
  80   & 557  & 160 &   13.5  &  0.092   \\
  96   & 438  & 280 &   16.1  &  0.077   \\
 \hline\hline
   64   & 609  &  160 &  6.03  &  0.206   \\
   \hline
\end{tabular}
\end{center}
\caption{Values of lattice size, $L$, number of measurements
$n_{meas}$ and number of gauge copies $n_{copy}$ used throughout
this paper. Spacing is $a=0.168$~fm. The data in the last line correspond
to $\beta=7.09$ ($a=0.094$~fm).
}
\label{tab:data_sets}
\end{table}

The features of the  gauge-fixing methods \cite{Bogolubsky:2007bw}
used in our study are as follows. Firstly, we extend the gauge group by the  transformations
(also referred to as $Z_2$ flips) defined as follows:
\begin{displaymath}
\label{def:nonperiodicGT}
 f_\nu(U_{x,\mu}) = \left[ \begin{array}{l}  -\; U_{x,\mu} \quad \mbox{if} \quad \ \mu=\nu \quad \mbox{and}\quad
x_{\mu} = a,   \\
 \ \ U_{x,\mu} \quad \mbox{otherwise}  \end{array} \right.
\end{displaymath}
which are the generators of the $Z_2^3$ group leaving the action (\ref{eq:action}) invariant.

Such flips are equivalent to nonperiodic gauge transformations.
A Polyakov loop directed along the transformed links
and averaged over the $2$-dimensional plane changes its sign.
Therefore, the flip operations combine
the $2^3$ distinct gauge orbits (or Polyakov loop sectors) of strictly
periodic gauge transformations into one larger gauge orbit.

The second feature is making use of the simulated annealing (SA),
which has been found computationally more efficient than the
use of the standard overrelaxation (OR) only \cite{Schemel:2006da,Bogolubsky:2007pq,
Bogolubsky:2007bw}.
The SA algorithm generates gauge transformations
$~g(x)~$ by MC iterations with a statistical weight proportional to
$~\exp{(3V~F_U[g]/T)}~$. The ``temperature'' $~T~$ is an auxiliary
parameter which is gradually decreased in order to maximize the
gauge functional $~F_U[g]~$.  In the beginning, $~T~$ has to be chosen
sufficiently large in order to allow traversing the configuration space
of $~g(x)~$ fields in large steps.
As in Ref. \cite{Bogolubsky:2007bw}, we choose $~T_{\rm init}=1.5$.
After each quasi-equilibrium sweep, including both heatbath and microcanonical updates,
$~T~$ is decreased with equal step size. The final SA temperature is fixed
such that during the consecutively applied OR algorithm the violation of
the transversality condition
\beq
{ga\over 2}\;\max_{x\mbox{,}\, a} \, \Big|
\sum_{\mu=1}^3 \left( A_{\mu}^a(x+{\hat \mu a\over 2}) - A_{\mu}^a(x-{\hat \mu a\over 2}) \right)
\Big| \, < \, \epsilon
\label{eq:gaugefixstop}
\eeq
decreases in a more or less monotonous manner for the majority of gauge fixing
trials until the condition (\ref{eq:gaugefixstop}) becomes satisfied with $\epsilon=10^{-7}$.
A monotonous OR behavior is reasonably satisfied for a final lower SA
temperature value $~T_{\rm final}=0.01~$~\cite{Schemel:2006da}.
The number of temperature steps is set equal to $3000$.
The finalizing OR algorithm using the standard Los Alamos type overrelaxation
with the parameter value $\omega = 1.7$ requires typically a number of
iterations of the order $O(10^3)$.

We then take the best copy out of many gauge fixed copies obtained for the
given gauge field configuration, i.e., a copy with the maximal value of the
lattice gauge fixing functional $F$ as a best estimator of the {\it global} extremum
of this functional.

In what follows, we consider three gauge-fixing methods.

The first one employs both the SA-OR algorithm and the $Z_2$ flips.
Using the SA-OR algorithm, we generate $8 n_c$ Gribov copies
($n_c$ copies in each $Z_2^3$ sector) and find the copy giving the
maximum of the functional (\ref{eq:gaugefunctional}). This copy is
referred to as the best  copy (``\bc'') and it should be mentioned
that we also find ``the best sector'' for each starting
configuration. The version of the Landau gauge  obtained by this
method is labelled FSA  (``Flipped Simulated Annealing'').

The second method employs the SA-OR algorithm, whereas
the $Z_2$ flips are not taken into consideration.
In this case, we choose ``the best copy'' from $n_c$
configurations, that corresponds to a random choice of the $Z_2^3$ sector.
We name it ``SA gauge-fixing'' (it is analogous
in some sense to the ``absolute gauge fixing'' in \cite{Maas:2008ri}).

In order to demonstrate the effect of Gribov copies,
we also consider the gauge obtained by a random choice
of a copy within the first Gribov horizon (labelled as ``\fc''---first copy).
That is, in the third case, we take the first copy obtained by the SA
algorithm and do not take care of fundamental modular domain at all.
This version of the Landau gauge is analogous in some sense to the ``minimal
Landau gauge'' in \cite{Maas:2008ri}).

Information on our simulation procedure is also shown in
Table~\ref{tab:data_sets}.  The scale is set by the relation
$a\sqrt{\sigma}\approx 0.376$ for $\beta = 4.24$ (see formula (65) in \cite{Teper:1998te}),
where $\sigma\approx (440$~MeV$)^2$ is the string tension
(which is assumed the same as in the four dimensions).

\section{Gribov copy effects and large $L$ behavior of $D(0)$}
\label{sec:gribov}

\begin{figure}[tb]
\centering
\includegraphics[width=7.1cm,angle=270]{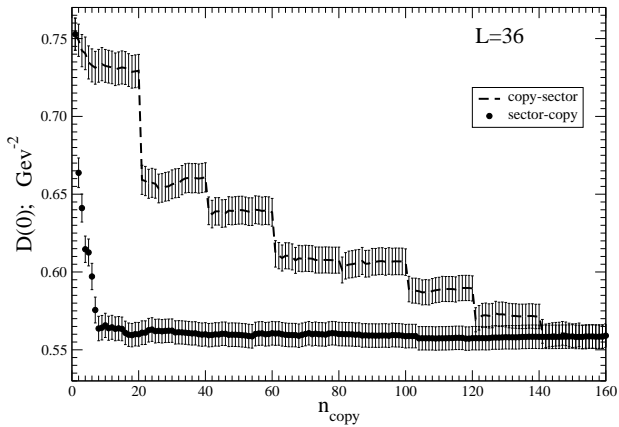}
\caption{$D(0)$ as a function of $n_{copy}$.
The meaning of 'copy-sector' and 'sector-copy' is explained
in the text.
}
\label{fig:D0_vs_nc}
\end{figure}

To demonstrate the Gribov copy effects, we show in \Fig{fig:D0_vs_nc}
the dependence of $D(0)$ on the number of gauge copies $n_{copy}$ for
$L=36$ lattice. As one can see, the Gribov copy influence is very strong
(at least, for $p=0$), and by no means can be considered as a "Gribov
noise".

\begin{figure}[tb]
\centering
\includegraphics[width=7.1cm,angle=270]{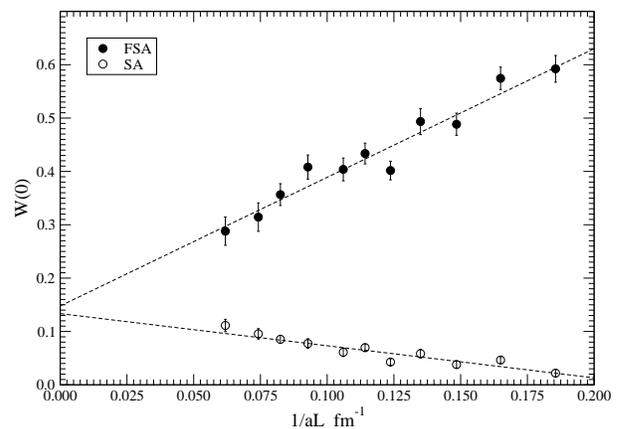}
\caption{$W(0)$ as a function of $1/aL$ for FSA and SA.
}
\label{fig:W0_vs_L}
\end{figure}

For the upper curve in this Figure ('copy-sector') first 20 gauge
copies belong to the first flip-sector, next 20 gauge copies
belong to the second flip-sector, etc..  Evidently, the dependence
on the number of copies belonging to the same flip-sector is
rather weak (apart from the first one), and the main Gribov copy
effect comes from different flip-sectors. This is demonstrated
also by the lower curve ('sector-copy')
where we have used another enumeration of gauge copies, i.e.,
$n_{copy}=1$ corresponds to the 1st copy of the 1st flip-sector,
$n_{copy}=2$ means that first copies of the first and second
flip-sectors are considered, etc..
With increasing volume $n_{copy} -$ dependence is changing as
we will explain  below in discussion of \Fig{fig:W0_vs_L}.

\vspace{1mm}

To compare Gribov copy effects for different values of $L$ and $p\ne
0$, we define the Gribov copy sensitivity parameter $\Delta(p) \equiv
\Delta(p;L)$ as a normalized difference of the \fc and \bc gluon
propagators

\beq
\Delta(p) = \frac{D^{fc}(p) - D^{bc}(p)}{D^{bc}(p)}~,
\label{delta}
\eeq
\noi where the numerator is the average of the differences
between \fc and \bc propagators calculated for every configuration
and normalized with the \bc (averaged) propagator.

In \Tab{tab:gceFSAgauge} we show the values of the parameter $\Delta(p)$
for different values of $L$ and four momenta : $p=0$, $p_{min}$,
$\sqrt{2}\,p_{min}$ and $\sqrt{3}\,p_{min}$. It is interesting to note
that values of this parameter do {\it not} demonstrate the tendency
to decreasing with increasing lattice size $L$ for all momenta under
consideration. In particular, at zero momentum the Gribov copy effect
is estimated to be between appr.  $25\%$ and $35\%$ (taking into account
the error bars) for all values of $L$.


\begin{table}[h]
\begin{center}
\vspace*{0.2cm}
\begin{tabular}{|c|c|c|c|c|} \hline
$L$ &  $p=0$ & $\,p_{min}$ & $\sqrt{2}\,p_{min}$ & $\sqrt{3}\,p_{min}$  \\ \hline\hline
 32 &  $0.318(14) $ & $0.107(7) $  & $0.043(5)  $ & $ 0.014(6) $  \\
 36 &  $0.329(13) $ & $0.107(6) $  & $0.039(4)  $ & $ 0.013(6) $  \\
 40 &  $0.289(13) $ & $0.104(7) $  & $0.039(5)  $ & $ 0.031(6) $  \\
 44 &  $0.308(16) $ & $0.100(8) $  & $0.042(5)  $ & $ 0.026(7) $  \\
 48 &  $0.251(11) $ & $0.100(5) $  & $0.046(4)  $ & $ 0.029(5) $  \\
 52 &  $0.285(13) $ & $0.091(6) $  & $0.043(4)  $ & $ 0.031(6) $  \\
 56 &  $0.273(15) $ & $0.116(8) $  & $0.042(5)  $ & $ 0.023(7) $  \\
 64 &  $0.294(17) $ & $0.100(8) $  & $0.049(6)  $ & $ 0.027(8) $  \\
 72 &  $0.261(16) $ & $0.121(7) $  & $0.052(5)  $ & $ 0.044(6) $  \\
 80 &  $0.232(20) $ & $0.102(10) $ & $0.062(7)  $ & $ 0.031(8) $  \\
 96 &  $0.218(21) $ & $0.115(12) $ & $0.074(8)  $ & $ 0.055(9) $  \\
 \hline\hline
 64 &  $0.308(17) $ & $0.110(8) $  & $0.048(6)  $ & $ 0.016(8) $  \\
 \hline
\end{tabular}
\end{center}
\caption{Values of $\Delta(p)$  for the  FSA Landau gauge propagators
at $p=0$, $p=p_{min}$, $p=\sqrt{2}\,p_{min}$ and $p=\sqrt{3}\,p_{min}$.
The data in the last line correspond to $\beta=7.09$ ($a=0.094$~fm).
}
\label{tab:gceFSAgauge}
\end{table}

Therewith, for a given value of $L$, the parameter $\Delta(p)$ decreases quickly
with an increase of the momentum.
This observation is in agreement with the observations made earlier for
the four-dimensional $SU(2)$ theory \cite{Bornyakov:2009ug}.

In the last line of the Table~\ref{tab:gceFSAgauge} we present results for
two times smaller lattice spacing obtained on lattices with  $L=64$ at $\beta=7.09$.
This lattice matches in physical size the lattice $L=36$ at $\beta=4.24$.
As can be seen from comparison between the first and the last lines of the table
our data indicate that the strength of the Gribov copies effects does not depend
on the lattice spacing. This observation agrees with our earlier results obtained in
4D SU(2) gluodynamics \cite{Bornyakov:2009ug}.

\vspace{2mm}

We have attempted to estimate the infinite-volume limit of the
zero-momentum propagator $D(0;L)$, i.e., the limit $L\to\infty$.

One can apply various fit-formulas for this purpose, e.g., $~c_1+c_2/L$;
$~c_2/L^{c_3}$; $~c_1+c_2/L^{c_3}$, etc., and many of them
fit nicely {\it if} the values of $L$ are not very large (as in our case). However,
calculations \cite{Cucchieri:2007md} on the lattice with $L=320$ and
$\beta=3.0$ have shown that the first fit-formula is supposed to be
the preferable one (at least, in the minimal Landau gauge).  Therefore,
following \cite{Cucchieri:2007md} we apply the fit-formula
\beq
D(0;L) =  c_1 + c_2/L~
\label{eq:fit}
\eeq

\noindent to determine the $L\to \infty$ limit of $D(0;L)$.

\begin{figure}[tb]
\centering
\includegraphics[width=7.1cm,angle=270]{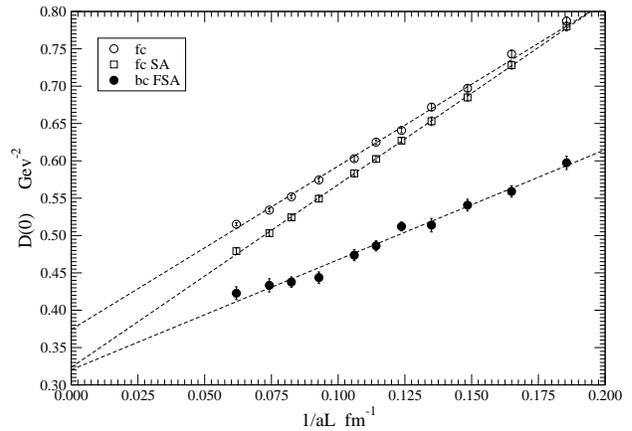}
\caption{$D(0)$ as a function of $1/aL$ for \fc ( the same for SA and FSA), \bc SA
(without flips) and \bc FSA.
}
\label{fig:D0_vs_L}
\end{figure}

In \Fig{fig:D0_vs_L} we show our values of $D(0;L)$ calculated for {\rm
a)} \fc (which is the same for SA and FSA methods); {\rm b)} \bc SA method
(i.e., without flips) and {\rm c)} \bc FSA method. Broken lines represent
fits according to \Eq{eq:fit}.

We confirm that in the $L\to\infty$ limit the value of $D(0)$ {\it
differs} from zero.  This is in agreement with the statement
made in \cite{Cucchieri:2007md} and is not in conformity with
\cite{Zwanziger:2003cf} and \cite{Maas:2008ri}.

\Fig{fig:D0_vs_L} shows another interesting phenomenon~: the Gribov copy
influence survives even in the thermodynamic limit $L\to\infty$. Indeed,
the infinite volume-extrapolation of $D^{fc}(0)$ {\it differs} from
infinite-volume extrapolation of $D^{bc}(0)$ (both for SA and FSA gauge
fixing algorithms).

To illustrate this phenomenon in a more explicit way, we calculated also
the averaged (over all configurations) difference between \fc and \bc
propagators normalized to $D^{bc}(p;L=\infty)$
\begin{equation}
W(p) = \frac{D^{fc}(p) - D^{bc}(p)}{D^{bc}(p;L=\infty)}
\label{eq:W}
\end{equation}
In \Fig{fig:W0_vs_L} we show the dependence
of $W(0)$ on the inverse lattice size both for SA and FSA algorithms.
One can see that both algorithms predict {\it nonzero} difference
between \fc and \bc values of the propagators, this difference
being rather big ($\sim 15\%$) even in the thermodynamic limit.
Note that both SA and FSA algorithms give the coincident
(within errorbars) results in the $L\to \infty$ limit.

For small volumes $W(0)$ for SA algorithm, $W_{SA}(0)$,  is close to zero
while $W_{FSA}(0)$  is at its maximum. This corresponds
to strong effects of flip-sectors seen in Fig.~\ref{fig:D0_vs_nc}.
With increasing volume $W_{SA}(0)$ is increasing
indicating the increasing role of the copies within given
flip-sector. In opposite, decreasing of $W_{FSA}(0)$
with increasing volume implies that the role of the flip-sectors
reduces. In the infinite volume limit the two curves in
\Fig{fig:W0_vs_L} converge. This means that in this limit one
randomly chosen flip-sector is sufficient or, in other words, all
flip sectors are equivalent. Note that on our largest volume the
effect of the flip-sectors is still dominating over the effect of
the copies within one sector.

\begin{figure}[tb]
\centering
\includegraphics[width=7.1cm,angle=270]{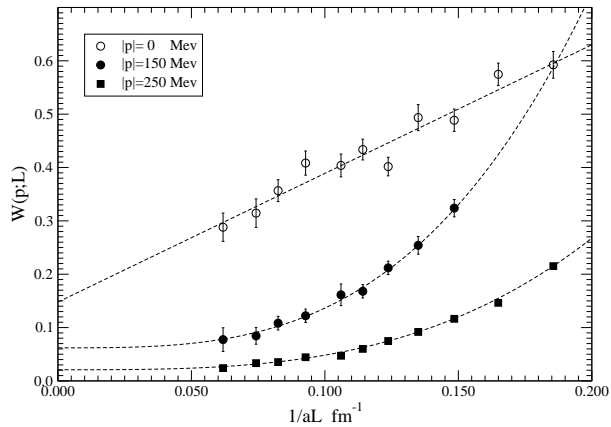}
\caption{$W(p)$ as a function of $1/aL$ for FSA
for three different values of $|p|$.
}
\label{fig:Wp_vs_L}
\end{figure}

In \Fig{fig:Wp_vs_L} we compare the $L$-dependence of $W(0)$ shown
in the previous Figure with that for two nonzero values of momenta:
$|p|=150$ MeV and $|p|=250$ MeV (all calculated for FSA).
The respective values of the propagators for fixed physical
momenta were obtained by interpolation of the original data to
necessary value of momentum. As it was expected, Gribov copy
effects decrease quickly with increasing $|p|$. However, they
are expected to be {\it not} small in the deep infrared region, i.e.,
in the $|p|\to 0$ limit. This can be essential for the calculation
of, say, screening masses in $4d$ theory at nonzero temperature
where the IR-behavior of the gluon propagator is important.

\vspace{2mm}

\begin{figure}[tb]
\centering
\includegraphics[width=7.1cm,angle=270]{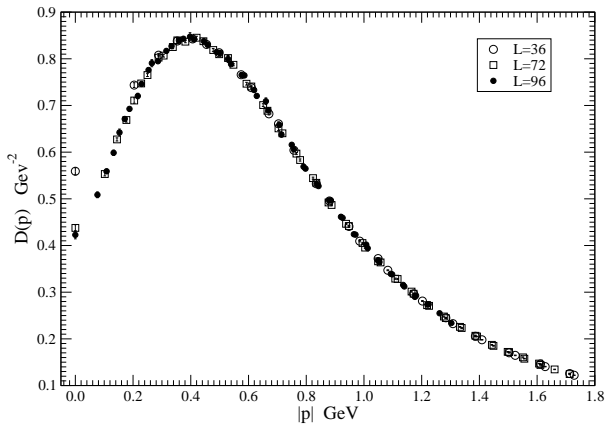}
\caption{The momentum dependence of $D(p)$ for three
lattices. }
\label{fig:D0_vs_p}
\end{figure}

In \Fig{fig:D0_vs_p} we compare the momentum dependence of the
propagators $D(p)$ calculated on $36^3$, $72^3$ and $96^3$ lattices.
Qualitatively its momentum dependence agrees with that obtained earlier
\cite{Maas:2006qw,Maas:2008ri,Cucchieri:2011ig}. But in contrast to
these papers the finite-volume effects are very small and can safely
be neglected at momenta $p>0.3$~GeV even for our smallest lattice.
Comparing results for one flip sector with results for 8 flip sectors
we conclude that the use of all flip sectors gives rise to drastic
decreasing of the finite-volume effects at momenta $p < 500$ MeV.

Note that the propagator has a maximum at nonzero value of
momentum  $|p| \sim 400$~MeV. Therefore, the behavior of
$D(p)$ in the deep infrared region is inconsistent with a simple
pole-type dependence.

\vspace{1mm}

\section{Conclusions}
\label{sec:conclusions}

In this work we investigated numerically the Landau gauge gluon propagator
$D(p)$ in the  three-dimensional pure gauge  $SU(2)$ lattice theory.
We have employed eleven lattice volumes from $L=32$ to $L=96$ at $\beta=4.24$
($a = 0.168$ fm) and one lattice volume $L=64$ at $\beta=7.09$ ($a=0.94$ fm).

One of our main goals was to study the approach of the zero-momentum
propagator $D(0)$ to the thermodynamic limit $L\to \infty$.
Special attention in this study has been paid to the dependence on the
choice of Gribov copies. To this purpose we have generated up to 280 gauge
copies for every configuration. Our \bc FSA method provides systematically
higher values of the gauge fixing functional as compared to the \fc FSA and \bc SA methods.
We conclude that the choice of the efficient gauge fixing procedure
is of crucial importance in the  study of the gluon propagator in the Landau gauge.
\vspace{1mm}

The main results are the following.

\begin{itemize}

\item[{\bf 1.}] As it can be seen from Fig.~\ref{fig:D0_vs_L} in the
limit $L\to\infty$ the value of $D(0;L)$ differs from zero. This is in
agreement with conclusions of Ref.~\cite{Cucchieri:2007md}.

\item[{\bf 2.}] We found that the Gribov copy influence is very strong in the deep
infrared region (especially for $D(0)$), and by no means can be
considered as a "Gribov noise".

Moreover, our analysis shows that the Gribov-copy effects remain
substantial  (up to $\sim 15 \%$ for the zero-momentum propagator)
even in the thermodynamic limit $L\to\infty$.

\item[{\bf 3.}]
We observed in agreement with earlier results in four-dimensional theory
\cite{Bogolubsky:2007bw}
that the use of the flip sectors dramatically decreased
finite-volume effects at momenta smaller than 500 MeV. Furthermore,
if all flip sectors are taken into account then finite-volume effects
for the gluon propagator at momenta greater than 300 MeV are negligibly
small even on our smallest lattice.

\item[{\bf 4.}]
The comparison of our results for the Gribov copy effects obtained
on two matched lattices with different lattice spacings (0.168 fm and
0.094 fm) indicates that the Gribov copy effects for $D(p)$
depend weakly (if any) on the lattice spacing.

 \end{itemize}

Our results are mainly obtained for rather large lattice spacing, thus
they should be further verified in the continuum limit.  We are now
doing such verification, results of the ongoing study will be published
elsewhere~\footnote{V.G. Bornyakov, V.K. Mitrjushkin, R.N. Rogalyov,
in preparation.}.

\vspace{2mm}

\subsection*{Acknowledgments}
This investigation has been partly supported by the Heisenberg-Landau
program of collaboration between the Bogoliubov Laboratory of Theoretical
Physics of the Joint Institute for Nuclear Research Dubna (Russia)
and German institutes, by the connected grants DFG Mu 932/7-1 and
RFBR 11-02-91339-NNIOa,  by  the Federal Special-Purpose
Programme "Cadres" of the Russian Ministry of Science and Education. VB
is supported  by grant RFBR 11-02-01227-a.

\bibliographystyle{apsrev}
\bibliography{citations_3d}

\end{document}